\documentclass[preprint,3p,times,a4paper,12pt]{elsarticle}
\usepackage{amssymb}
\usepackage{amsmath}
\usepackage{graphicx}
\usepackage{dcolumn}
\usepackage{bm}
\usepackage{color}

\journal{Nuclear Physics A}

\newcommand{\be}{\begin{equation}}
\newcommand{\ee}{\end{equation}}
\newcommand{\bea}{\begin{eqnarray}}
\newcommand{\eea}{\end{eqnarray}}

\newcommand{\bfq}{\mbox{\boldmath $q$}}

\newcommand{\bfp}{\mbox{\boldmath $p$}}

\newcommand{\bftau}{\mbox{\boldmath $\tau$}}

\newcommand{\bfsigma}{\mbox{\boldmath $\sigma$}}

\newcommand{\PGR}[1]{{\color{blue} #1}}

\begin{document}

\begin{frontmatter}

\title{Landau-Migdal vs. Skyrme}

\author[Juel,Bonn]{J. Speth}
\author[Juel,Bonn]{S. Krewald}
\author[Juel]{F. Gr\"ummer}
\author[Erl]{P.-G. Reinhard}
\author[StP]{N. Lyutorovich}
\author[StP]{V. Tselyaev}
\address[Juel]{Institut fur Kernphysik, Forschungszentrum J\"ulich, D-52425 J\"ulich, Germany}
\address[Bonn]{Inst. f. Kernphysik, Universit\"at Bonn,
Nussallee 10, Bonn /Germany}
\address[Erl]{Inst. f. Theor. Physik, Universit\"at Erlangen,
Staustr.7, D-91058 Erlangen /Germany}
\address[StP]{V.A. Fock Institute of Physics, St. Petersburg State University, RU-198504 St. Petersburg, Russia}

\begin{abstract} 
The magnitude and density-dependence of the non-spin dependent Landau-Migdal parameters are derived from Skyrme energy functionals and compared with the phenomenological ones. We perform RPA calculations with various approximations for the Landau-Migdal particle-hole interaction and compare them with the results obtained with the full Skyrme interaction. For the first time the next to leading order in the Landau-Migdal approach is considered in nuclear structure calculations.

\end{abstract}


\date{\today}

\end{frontmatter}

\section{Personal recollections by Josef Speth}

I met Gerry for the first time 1972 in Osaka at a conference on magnetic moments. He invited me to Copenhagen and Stony Brook where I spent my first sabbatical in 1975. From that time on we met several times a year in Stony Brook, in J\"ulich and at various conferences. He spent the time of his \emph{Humboldt award} in our institute in J\"ulich and became an expert for walking trails on the \emph{Sophien H\"ohe}, the artificial moment near J\"uelich. Last but not least I was his tennis partner and instructor. 

\section{Introduction}

Landau's \emph{Theory of Fermi Liquids} is a widely used and powerful
approach to describe excitation properties of extended Fermion system
\cite{LanLif9}. It has been generalized by Migdal to a \emph{Theory of
  Finite Fermi Systems} \cite{Migdal67}. Since the early 1970 G.E. Brown was interested in the Landau-Migdal approach. In the famous article \emph{Landau,Brueckner-Bethe and Migdal Theories of Fermi Systems} \cite{GE1} he reviewed and compared the most successful many-body approaches of that time. In connection with the pion condensation first discussed by Migdal, he pointed out \cite{GE2} that if one considers in Migdals calculation the spin-isospin dependent zero-range parameter $g^{\prime}{_0}$ of the Landau-Migdal interaction the condensation disappears. In the same year Babu and Brown \cite{GE3} studied the quasi-particle interaction in $^3$He where they introduced the \emph{induced interaction}, which allowed to satisfy the Pauli principle. Some years later he showed that in the spin-isospin channel of the nuclear particle-hole interaction one has to consider the pion and rho exchange contribution \cite{GE4}. This \emph{Stony-Brook J\"ulich interaction} was successfully applied a large body of magnetic properties . \\
In the 1990's, R. Shankar pointed out that the application of the renormalization group to
rotationally invariant Fermi surfaces automatically leads to Landau's Fermi-liquid theory as
a fixed point of the renormalization group flow \cite{Shankar:1993pf}. This observation was transferred from 
condensed matter physics to nuclear physics.  Renormalization group techniques led to the derivation of the unique $V_{low k}$ two-nucleon interaction in the vacuum. In a series of papers, Brown, Schwenk, and Friman applied renormalization group methods to derive the nuclear Fermi-liquid theory starting from the   $V_{low k}$ two-nucleon interaction
and the Babu-Brown induced interaction \cite{Schwenk:2001hg,Schwenk:2002fq}. 
 These studies opened a new approach first to
neutron matter, and eventually to nuclear matter and finite nuclei. A summary of the
present status of this field can be found in \cite{Friman:2012ft}.

In the following, we want to concentrate on another extension of the Landau theory,
based on effective interactions defined in a nuclear medium, the so-called Skyrme forces.
These interactions have a simple mathematical structure which has facilitated extensions and
applications of the
theory beyond the mean-field approximation.

	Landau-Migdal theory is based formally on a low-$q$ (low momentum) expansion of the effective
two-body interaction in the medium while the (few) model parameters
are adjusted phenomenologically.  The single-particle basis for the
RPA (random-phase approximation) calculation with the Landau-Migdal
interaction is taken from an empirical shell-model potential. The
approach was taken up and further developed in detail for nuclear
physics by the M\"unchen-J\"ulich group \cite{Zamick74}. Since then it
has been applied extensively for a broad range of nuclei, for a review
see \cite{Rev77}. 
At about the same time, another effective interaction from a low-$q$
expansion appeared in nuclear physics, the Skyrme-Hartree-Fock (SHF)
approach \cite{Sky59a,Neg72a,Neg75a,Liu76}. It was constructed with
different intention, predominantly as a self-consistent model for the
nuclear ground state\cite{Brink72}, but also applicable to compute
excitation spectra within RPA \cite{Kre77a}. For a review on SHF see
\cite{Ben03aR}.
The formal similarity of both approaches raises interest in a closer
comparison. Thus the microscopic calculation of the phenomenological
Landau- Migdal parameters from Skyrme energy functionals has a long
history \cite{Jackson75}. As self-consistent calculations are
considered to be more fundamental then the Landau-Migdal approach,
differences between the calculated and phenomenological parameters
were assumed as a short coming of the Landau-Migdal theory. That this
is not necessarily correct showed the former discussion on the  incompressibility $K$ 
and the excitation energy of the breathing
mode in $^{208}$Pb which both are related to the parameter
F$_{0}^{in}$. The value from Landau-Migdal theory of the
M\"unchen-J\"ulich group \cite{Zamick74,Rev77} was nearly zero, which
strongly deviated from values derived from the early SHF
parametrizations \cite{Brink72}. The value of the incompressibility
of the order of $K\approx 250$ MeV and the predicted energy of the
breathing mode deduced from the Landau-Migdal theory turned out to be
close to data, whereas the old Skyrme value of the incompressibility
$K\approx 350 MeV$ was much too high. It was found in the next
  stage of SHF development that this deficiency was due to a too rigid
  modeling of the density dependence of the Skyrme interaction and
  that a more flexible density dependence can remove the discrepancy
  \cite{Bar82a,Bra85aR}. This little historical example shows that a
  comparison of these two similar and yet different low-$q$ models
  can be fruitful. It is the aim of this paper to continue those
  comparison at an up-to-date level of modeling.

In the present comparison, we will address two aspects: first, the
  prediction and density-dependence of Landau-Migdal parameters
  (LMP) modern SHF models, and second, the impact of
  kinetic terms in RPA calculations of nuclear giant resonances.  In
Landau's theory, the interaction parameters are constants. In Migdal's
extension, the force parameters are density-dependent to account
  for the finite size effects in the nuclei. SHF predicts a
  density dependence for the LMP. We will compare the different
  density dependence and we will address the uncertainties in the SHF
  predictions on the basis of the techniques of error propagation in
  least-squares fits \cite{Klu09,Kor10}.  For the studies of RPA
  excitations, we will employ a recently developed an RPA code which
  can perform Calculations within Landau-Migdal theory as well as
  with the full SHF residual interaction. This allows a direct comparison on the basis of
  the same numerical treatment. We use that for studying the impact of
the kinetic terms and the Landau approximation.

The paper is outlines as follows:
Section \ref{sec:LMT} provides a brief review of Landau-Migdal theory,
section \ref{sec:SHFpred} discusses the SHF predictions of LMP,
and section \ref{sec:RPA} contains a study of the impact of
kinetic terms on RPA spectra.

\section{Landau-Migdal Theory}
\label{sec:LMT}

The central object of the Landau-Migdal approach \cite{Migdal67} is the response function $L$ which is defined as:
\begin{equation}\label{eq:4}
L(13,24)=g(13,24)-g(1,2)g(3,4).   
\end{equation}
where $g(1,2)$  and $g(13,24)$ are the one-body and two body Green functions. The response function obeys an integral equation: 
\begin{equation}\label{eq:5}
L(13,24)= -g(1,4)g(3,2) - i\int{d5d6d7d8g(1,5)K(57,68)L(83,74)g(6,2)}
\end{equation}
where $K$ is an effective two-body interaction. It is irreducible with respect to the particle-hole propagator.
In Eq.(\ref{eq:5}) the kernel $K$ as well as the one-particle Green
functions are a priori unknown. The one-particle Green function is given by the
Dyson equation: 
\begin{subequations}
\begin{eqnarray}
\label{eq:24}
  &
  \frac{i}{2}\int{d3\left\{S(1,3)\;+\;\Sigma{(1,3)}\right\}g{(3,2)}}
  =
  \delta(1,2)
  \quad,
\\
\label{eq:23}
  & 
  S(1,3) 
  = 
  \delta_{\nu_1\nu_2}\delta(t_1-t_3)
  \left\{i\frac{\delta}{\delta{t_1}}-\epsilon_{\nu_1}^0\right\} 
  \quad,
\end{eqnarray}
\end{subequations}
where $\Sigma{(1,3)}$ is the mass operator which is connected in a
very complicated way with the two- body interaction of a given
Hamiltonian \cite{Rev77}.  The irreducible kernel $K$ is defined by
the functional derivative \cite{Rev77} :
\begin{equation}\label{eq:27}
K = \frac{\delta \Sigma (1,2)}{\delta g(3,4)}. 
\end{equation}

 In general, the effective interaction kernel is an involved four-point
function. In momentum representation, it is a function of four momenta,
however the energy- and density-independent part of $K$
(in particular, the density-independent terms of the Skyrme interaction)
depends actually only on three momenta due to the translation symmetry.
For this part we have $K=K(\mathbf{p},\mathbf{p}',\mathbf{q})$ where
$\mathbf{p}$ and $\mathbf{p'}$ are the momenta of the in-coming and
out-going hole states and $\mathbf{q}$ is the transferred momentum. $K$ is also energy dependent. 
\begin{figure}[htbp]
\begin{center}
\includegraphics[width=0.7\linewidth]{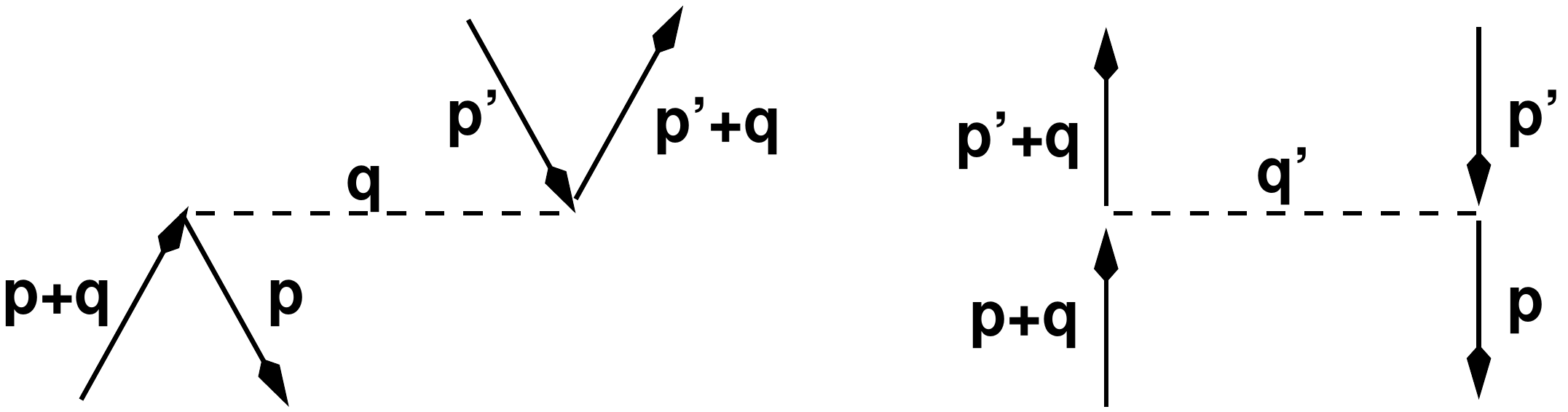}
\end{center}
\caption{\label{fig:basicV}
Graphical representation of a local interaction in $ph$ space
with direct (left) and exchange term (right). Particle and
hole states are represented by full lines with up- and down arrows. 
The dashed horizontal line stands for the interaction.
The three relevant momenta $\mathbf{p}$,  $\mathbf{p}'$, and
$\mathbf{q}$, are indicated. The exchange term transfers the
momentum  $\mathbf{q}'=\mathbf{p}'-\mathbf{p}$.
}
\end{figure}
Figure \ref{fig:basicV} illustrates these momenta for the case of a
local interaction which is sufficient for our purposes because the
Landau-Migdal as well as the Skyrme interaction are both local.
All $ph$ pairs carry net momentum $\mathbf{q}$. They differ by the
other momentum $\mathbf{p}$, or $\mathbf{p}'$ respectively. 

Following the quasi-particle concept of Landau, one separates the
one-particle Green function into a singular quasi-particle term and a
remainder.  With this ansatz one can rewrite Eq.(\ref{eq:5}) and
obtains after some analytical transformations (see Ref. \cite{Rev77})
the renormalized RPA equation for nuclei in terms of the excitation
amplitudes $\chi$ in the single-particle configuration space
\cite{Ring80}:
\begin{equation}\label{eq:9a}
\left( \epsilon_{\nu_1}-\epsilon_{\nu_2}- \Omega_m\right)\chi^{(m)}_{\nu_1 \nu_2} =  
\left(n_{\nu_1}-n_{\nu_2}\right)\sum_{\nu_3 \nu_4}\rm{F^{ph}}_{\nu_1 \nu_4 \nu_2 \nu_3} \chi^{(m)}_{\nu_3 \nu_4}.
\end{equation}
F$^{ph}$ is the ph-interaction, $\Omega_m$ are the excitation energies
of the nucleus and $\chi ^{m}$ the corresponding quasi-particle
quasi-hole transition matrix elements.  F$^{ph}$ is a complicated
function of $K$ and the non-singular parts of the Green functions
e.g. $K$ appears in the nominator as well as in the denominator
\cite{Rev77}. Therefore F$^{ph}$ is a smooth function in momentum
space and correspondingly of short-range in the $\mathbf{r}$-space.

Moreover, in the Landau approach one considers the interaction on the
Fermi surface and replaces the energies by the Fermi energy and the
magnitude of the momenta by the Fermi momentum. Thus one can
approximate F$^{ph}$ as a local contact (zero-range) interaction. This
means that F$^{ph}$ is effectively independent on $\mathbf{q}$ and
$\mathbf{q'}$, respectively. After all, F$^{ph}$ depends only on the
angle between the ph-momenta $\mathbf{p}$ and $\mathbf{p'}$ before and
after the collision; it reads
\begin{equation}\label{eq:3a}
  F^{ph}\left( \frac{ \mathbf{p}\cdot \mathbf{p'}} {p^{2}_F} \right) 
  = 
  C_0\sum_{l=0}^{\infty} 
  \left[f_l+f^{\prime}_l\hat{\mathbf{\tau_1}}\cdot\hat{\mathbf{\tau_2}} 
       +g_l\hat{\mathbf{\sigma_1}}\cdot\hat{\mathbf{\sigma_2}}
       +g^{\prime}_l\hat{\mathbf{\sigma_1}}\cdot\hat{\mathbf{\sigma_2}}
                   \hat{\mathbf{\tau_1}}\cdot\hat{\mathbf{\tau_2}}\right]
      P_l\left( \frac{ \mathbf{p}\cdot \mathbf{p'}} {p^{2}_F} \right)
\end{equation}
where $P_l(x)$ is the Legendre polynomial of order $l$ and the four
terms containing different combinations of spin and isospin operators
cover the typical four nuclear interaction channels \cite{Ring80}.
Note the remarkable result that, by virtue of the Landau
quasi-particle concept and the following renormalization, the whole
information content of $K$ shrinks to a few model constants, the much
celebrated Landau-Migdal parameters $f_l$, usually restricted to $l=0$
and 1. These parameters are dimensionless and $C_0$ is defined as:
\begin{equation}\label{eq:4a}
  C_0 = \frac{\pi^2 \hbar^2}{2 m^* k_{\rm F} } 
  \end{equation}
where $k_\mathrm{F}=\left(3\pi^2\rho_0/2\right)^{1/3}$ is the Fermi
momentum. The scaling factor $C_0$ is proportional to the 
density of states at the Fermi surface.  A typical value is
$C_0=150\,\mathrm{MeV}\,\mathrm{fm}^3$ which is the standard
choice in phenomenological shell models where the effective mass is
$m^*/m=1$.  It is to be noted that papers from the Landau-Migdal
theory often use a factor which is twice as large \cite{Migdal67}. 
The above scaling (in its flexible form with actual
$m^*$ and $k_\mathrm{F}$) is the standard in all SHF papers addressing
Landau parameters. We will follow this option henceforth.

The Fourier transforms of the ($\mathbf{q}$-independent) terms with
$l=0$ and $l=1$ yield $\delta$-functions and derivatives of
$\delta$-functions in coordinate space, of the above mentioned zero-range
interactions.

In the \emph{ Theory of Fermi Liquids} the Landau parameters
are constants. Migdal introduced in his \emph{Theory of Finite Fermi
  Systems} density dependent parameters $f_l(\rho)$ in order to 
correct for the finite size of the nuclei.  The form of the
interaction in leading order ($l=0$) in the $\mathbf{r}$-space is thus
written as:
\begin{equation}\label{eq:5a}
F^{ph}(1,2)=C_{0}\delta (\mathbf{r_1}-\mathbf{r_2})\cdot 
\left[f_0(\rho) + f^{\prime}_0(\rho)\mathbf{\tau_1} \cdot\mathbf{\tau_2} +
      g_0(\rho)\mathbf{\sigma_1} \cdot\mathbf{\sigma_2} \\
      +g^{\prime}_0(\rho)
      \mathbf{\sigma_1}\cdot\mathbf{\sigma_2}\mathbf{\tau_1}
      \cdot\mathbf{\tau_2}\right]
\;. 
\end{equation}
The density dependent Landau-Migdal parameters are parametrized in the
following way \cite{Migdal67}:
\begin{equation}\label{eq:22}
  f(\rho) 
  = 
  f^\mathrm{(ex)} + (f^\mathrm{(in)}-f^\mathrm{(ex)})\frac{\rho_0(r)}{\rho_0(0)}
\end{equation}
where $f^\mathrm{(ex)}$ stands for the exterior region of the nucleus
and $f^\mathrm{(in)}$ for the interior.  So far only the leading-order
contribution of the Landau-Migdal interaction has been considered in
nuclear structure calculations. There is a similar expansion for the
spin-spin part of the effective interaction leading to the
Landau-Migdal parameters $g_l$, or $g^{\prime}_l$ respectively for the spin
response. We will not address spin response in this paper.

The question remains how to determine the model parameters.  In the
Landau-Migdal approach neither, $g(1,2)$ nor $K$ are derived
microscopically from Eqs. (\ref{eq:5}--\ref{eq:27}) but are obtained
by adjustment to phenomenological data.  The parameters $f_0^\mathrm{(in)}$ and 
$f{^{\prime}}^\mathrm{(in)}_0$ 
are related to the compressibility and symmetry energy and
deduced from Eq.(\ref{eq:LandauNMP}). The external parameters were
adjusted to experimental data e.g. Ref. \cite{Rev2004}. 

In order to solve the basic Landau Migdal equations one needs as input
single particle-wave functions, single-particle energies and the
ph-interaction. Migdal has designed his theory in close connection to
Landau's \emph{ Theory of Fermi Liquids}.  Therefore one takes the
input data from experiment or from models which reproduce the needed
experimental data as good as possible. The single-particle wave
functions are taken from an empirical single-particle model and the
single-particle energies as far as possible from experiment.

\section{Landau parameters from the Skyrme energy-density functional}
\label{sec:SHFpred}

The original formulation of the Skyrme-Hartree-Fock (SHF) method was
based on the concept of an effective interaction, the Skyrme force
\cite{Sky59a}. It was observed long ago that the density dependence in
the Skyrme force inhibits an interpretation as interaction
\cite{Kre77a}. The theoretically correct attitude is to see the SHF
method as nuclear density functional approach. Many modern treatments
of SHF thus start from a Skyrme energy-density functional, see
e.g. the reviews \cite{Ben03aR,Erl11aR}. On the other hand, the Skyrme
force, being a zero-range interaction, has a great formal similarity
to the Landau-Migdal force. We thus use it here as the generator of
the Skyrme energy-density functional and write the functional
as expectation value 
\begin{subequations}
\label{eq:skenfun}
\begin{eqnarray}
  E_\mathrm{Sk}
  &=&
  E_\mathrm{Sk,dens}
  +
  E_\mathrm{Sk,grad}
  \quad =
\\
   \langle\Phi|\hat{V}_\mathrm{Sk}|\Phi\rangle 
\\
  E_\mathrm{Sk,dens}
  &=&
  \langle\Phi|
  t_0(1\!+\!x_0 \hat{ P}_\sigma)\delta(\mathbf{r}_{12})
  +
   \frac{t_3}{6}(1\!+\!x_3\hat P_\sigma)
  \rho^\alpha\left(\mathbf{r}_1\right)
  \delta(\mathbf{r}_{12})
  |\Phi\rangle
\nonumber\\
  E_\mathrm{Sk,grad}
  &=&
  \langle\Phi|
   \frac{t_1}{2}(1\!+\!x_1\hat{P}_\sigma)
  \left(
   \delta(\mathbf{r}_{12})\hat{\boldsymbol k}^2
   +
   {\hat{\boldsymbol{k}}}'^{2}\delta(\mathbf{r}_{12})
  \right)
  + t_2(1\!+\!x_2\hat P_\sigma)\hat{\boldsymbol k}'
  \delta(\mathbf{r}_{12})\hat{\boldsymbol k}
  |\Phi\rangle
\end{eqnarray}
\end{subequations}
where $\mathbf{r}_{12}=\mathbf{r}_1-\mathbf{r}_2$ and $\hat{P}_\sigma=
\frac{1}{2}(1+\hat{\boldsymbol \sigma}_1\hat{\boldsymbol{\sigma}}_2)$
is the spin-exchange operator. The momentum operators are
$\hat{\boldsymbol{k}}=-\frac{i}{2}\left(\stackrel{\rightarrow}{\boldsymbol\nabla}_1
-\stackrel{\rightarrow}{\boldsymbol\nabla}_2\right) $ and
$\hat{\boldsymbol{k}}' =
\frac{i}{2}\left(\stackrel{\leftarrow}{\boldsymbol\nabla}_1
-\stackrel{\leftarrow}{\boldsymbol\nabla}_2\right)$ where
$\hat{\boldsymbol k}$ acts to the right and $\hat{\boldsymbol{k}}'$ to
the left. 

Leading part is $E_\mathrm{Sk,dens}$ for which the interpretation as
energy-density functional is compulsory.  It can be expressed in terms
of local densities $\rho_p$, $\rho_n$ and spin densities
$\boldsymbol{\sigma}_p$, $\boldsymbol{\sigma}_n$, for details see
\cite{Ben03aR,Erl11aR}. 
The gradient part $E_\mathrm{Sk,grad}$ involves additionally
the gradients of density as well as spin-density and, as truly new
ingredients, kinetic-energy densities and currents. 
We ignore here the spin-orbit and the tensor contributions to the
Skyrme energy. They play no role for a comparison with Landau
parameters.

Writing the SHF energy in terms of the Skyrme force, as done above,
involves naturally the Skyrme parameters $t_i$ and $x_i$. Starting
from an energy-density functional suggest other parameter
combinations. To avoid confusion, it is preferable to express the SHF
functional in terms of nuclear matter properties (NMP) which have
one-to-one relation to the Skyrme parameters, except for the
spin-orbit and tensor part which is anyway not discussed here.
Consequently, it is most robust to express the Landau-Migdal
parameters as derived from the SHF functional in terms of NMP. As the
Landau-Migdal coefficients parametrize the response properties of a
system, it is natural that response parameters from nuclear matter
come into play, namely incompressibility $K$, symmetry energy
$a_\mathrm{sym}$, effective mass $m^*$, and isovector effective mass
also characterized by the Thomas-Reiche-Kuhn sum rule enhancement
$\kappa_\mathrm{TRK}$, for details see e.g. \cite{Erl11aR}. The
relations are

\begin{subequations}
\label{eq:LandauNMP}
\begin{eqnarray}
  K(\rho)
  =
  \frac{\hbar^2}{2m^*(\rho)}6k_\mathrm{F}^2(1+f_0(\rho))
  &\leftrightarrow&
  f_0(\rho)
  =
  \frac{2m^*(\rho)}{\hbar^2}\frac{K(\rho)}{6k_\mathrm{F}^2}-1
  \quad,
\\
  \frac{m^*(\rho)}{m}
  =
  1+\frac{f_1(\rho)}{3}
  &\leftrightarrow&
  f_1(\rho)=3\left(\frac{m^*(\rho)}{m}-1\right)
  \quad,
\\
  a_{sym}(\rho)
  = 
  \frac{1}{3}\frac{\hbar^2 k^2_F}{2m*(\rho)}(1+f'_0(\rho))
  &\leftrightarrow&
  f'_0(\rho)
  =
  \frac{2m^*(\rho)}{\hbar^2}\frac{3}{k_\mathrm{F}^2}a_\mathrm{sym}(\rho)-1
  \quad,
\\
  \kappa_\mathrm{TRK}(\rho)
  =
\frac{m}{3m^*(\rho)}\left(f'_1(\rho)-f_1(\rho)\right)
  &\leftrightarrow&
  f'_1(\rho)
  =
  3\left(\frac{m^*(\rho)}{m}\kappa_\mathrm{TRK}(\rho)-1\right)
  \quad.
\end{eqnarray}
\end{subequations}
These relations refer to the dimensionless Landau-Migdal parameters as
defined in Eq. (\ref{eq:4a}). Note that the lowest order coefficients
$f_o$ and $f'_0$ acquire an involved density dependence due to the
density dependence of the SHF functional $E_\mathrm{Sk,dens}$. Its
form differs from the simple linear interpolation Eq.  (\ref{eq:22}).

The SHF functionals as such are usually adjusted to empirical data
\cite{Ben03aR,Erl11aR}. However, once determined they constitute a
universal parametrization aiming at describing all nuclei (except the
smallest ones) as well as neutron and nuclear matter. They thus allow
to ``derive'' Landau-Migdal parameters with Eqs. (\ref{eq:LandauNMP}).
This is different from Landau-Migdal theory where the Landau-Migdal
parameters as such are adjusted phenomenologically. It is thus
interesting to compare the parameters derived from SHF with those from
pure Landau-Migdal theory. Scanning the SHF literature, one finds a
puzzling variety of predictions for the Landau-Migdal parameters which
is due to the fact that SHF fits with their weight on ground state
properties determine some aspects of the dynamical nuclear response
only loosely. A reliable protocol of these inherent uncertainties is
achieved by the rues of error propagation in connection with
least-squares fits \cite{Bevington}. Such systematic adjustment
studies are becoming increasingly fashionable in SHF studies
\cite{Klu09,Kor10}. We employ here two parametrizations, SV-min and
SV-bas, from \cite{Klu09} to demonstrate the impact of fit data on the
predictions of Landau-Migdal parameters. The parametrization SV-min
stems from a straightforward fit exclusively to ground state
properties (binding energies, radii, electro-magnetic formfactor) of a
large pool of finite nuclei. The extrapolation uncertainties of SV-min
are typical for all Skyrme forces fitted to ground state data.  The
parametrization SV-bas uses the same pool of data as SV-min, but
includes additionally information on response properties of
$^{208}$Pb, the isovector dipole polarizability and the peak energies
of three giant resonances: isoscalar monopole, isovector dipole, and
isoscalar quadrupole. These response data are similar to what is used
in fitting Landau-Migdal parameters and it is interesting to their
(indirect) impact through the SHF fits.

\begin{figure}[t]
\centerline{\includegraphics[width=\linewidth]{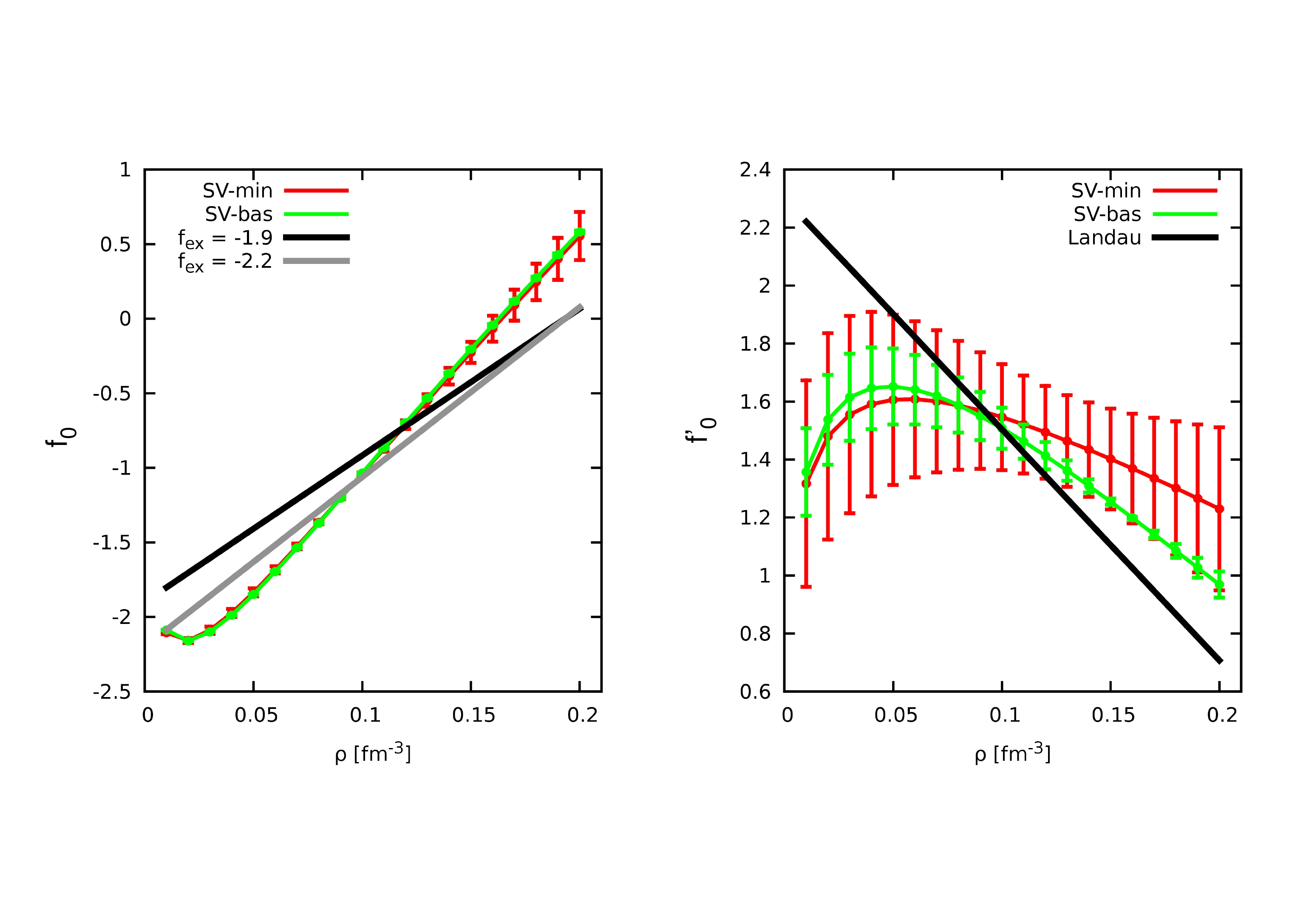}}
\caption{\label{fig:lanpar-uncert1} Dimensionless Landau-Migdal
  parameters in normalization (\ref{eq:4a}) for the Skyrme
  parametrizations SV-min and SV-bas together with the uncertainties
  from the $\chi^2$ fit which was employed to determine this
  parametrization \cite{Klu09}.  Additionally indicated are standard
  Landau-Migdal parameters (marked as ``Landau''). 
  }
\end{figure}
Figure \ref{fig:lanpar-uncert1} shows the dimensionless Landau-Migdal
parameters $f_0(\rho)$ and $f^\prime_0(\rho)$ for the SHF
parametrizations SV-min and SV-bas from \cite{Klu09}.  The
extrapolation uncertainties on the Landau-Migdal parameters are also
shown by error bars. The errors on $f_0$ are very small. At larger
densities, the errors increase with density and become visible for
SV-min. SV-bas has generally smaller errors because it includes more
data in the fit. Its errors remain below drawing precision for
$f_0$. The small errors for $f_0$ are plausible because isoscalar
properties are well determined by the the fits to known nuclei
\cite{Klu09,Erl11aR}.  The situation is much different for isovector
properties and accordingly we see large uncertainties for the
isovector parameter $f'_0$, particularly for SV-min. There is a slight
density dependence of its error. It is interesting to note that a
minimum of uncertainty for SV-min is found in the inner surface region
around densities $\rho\approx 0.12$ fm$^{-3}$.  This density
corresponds to the nuclear surface region to which the giant dipole
resonance is most sensitive.  The errors for $f'_0$ with SV-bas are
significantly smaller, particularly in the surface and volume region
($\rho\approx 0.1-0.16$ fm$^{-3}$). This demonstrates very clearly the
strong connection between response properties (which were included in
the fit of SV-bas) and Landau-Migdal parameters. It is interesting to
note that even for SV-bas the errors increase towards very low
densities. This outer surface region is not well determined, neither
by ground state data nor by resonance properties.

Figure \ref{fig:lanpar-uncert1} shows also the standard Landau-Migdal
parameters for comparison. These look at first glance much different
as their density dependence as given by Eq. (\ref{eq:5a}) is linear in
contrast to the much more involved density dependence of the SHF
results. However, just in the dynamically most relevant region
at surface densities $\rho\approx 0.1$ fm$^{-3}$ there emerges a nice
agreement between SHF prediction and empirical parameters.

The $l=1$ parameters f$_1$ and f$^\prime_1$ have a simple, linear
density dependence also in SHF. Thus we spare a figure showing all
trivial density dependence. We just quote the value at bulk
equilibrium density $\rho=0.16$ fm$^{-3}$. We find for SV-min
f$_1=-0.144\pm 0.215$ and f$'_1=0.071\pm 0.781$ , for
SV-bas  f$_1=-0.302\pm 0.000$  and f$'_1=0.778\pm 0.010$. 
These parameters are surprisingly little determined by ground state
data, but very well fixed by the response information from giant
resonances.  The empirical \PGR{LMP} are not well determined,
only in Ref. \cite{Migdal67} are numbers quoted: $\left|f_1\right|\approx$ 0,1-0,2 and $\left|f'_1\right|\leq $0,1.

\section{Landau approximation versus exact Skyrme-RPA}
\label{sec:RPA}

The residual interaction for RPA calculations of nuclear excitation
spectra can be deduced as first derivative of the mass operator, see
Eq. (\ref{eq:27}), or directly as second functional derivative of the
energy-density functional,
$K\equiv{F^{ph}}=\partial^2E_\mathrm{Sk}/\partial\hat{\rho}_1\partial\hat{\rho}_2$.
The purely density dependent part $E_\mathrm{Sk,dens}$ yields a
zero-range interaction in full compliance with the Landau-Migdal form
(\ref{eq:5a}). It is only the density dependence which differs from
the simple linear ansatz (\ref{eq:22}) as we have already seen in
figure \ref{fig:lanpar-uncert1}. More critical is the kinetic term
coming from the gradient functional $E_\mathrm{Sk,grad}$.  The correct
but tedious way to determine the effective $ph$ interaction
$F^{ph}_\mathrm{Sk,grad}$ goes through the second functional
derivative of $E_\mathrm{Sk,grad}$.  As this term corresponds to a
pure two-body interaction, we can evaluate $F^{ph}_\mathrm{Sk,grad}$
directly as the two-body matrix element between the four plane waves
with momenta  $\mathbf{p}$, ($\mathbf{p}+\mathbf{q})$, $\mathbf{p}'$, and
($\mathbf{p}'+\mathbf{q})$ as sketched in figure \ref{fig:basicV}. This
yields
\begin{subequations}
\begin{eqnarray}
 F^{ph}_\mathrm{Sk,grad}(\mathbf{p},\mathbf{p}',\mathbf{q})
 &=&
 \bigl[b^{(-)}_{00}1^{\sigma}1^{\tau} 
       +b^{(-)}_{10}(\bfsigma\bfsigma')\,1^{\tau}
       +b^{(-)}_{01}1^{\sigma}\,(\bftau\bftau')
       +b^{(-)}_{11}(\bfsigma\bfsigma')(\bftau\bftau')\bigr]
 \mathbf{q}^2
\nonumber\\
  &&
   \!\!\!+\bigl[b^{(+)}_{00}1^{\sigma}1^{\tau} 
       +b^{(+)}_{10}(\bfsigma\bfsigma')\,1^{\tau}
       +b^{(+)}_{01}1^{\sigma}(\bftau\bftau')
       +b^{(+)}_{11}(\bfsigma\bfsigma')(\bftau\bftau')\bigr]
 (\mathbf{p}\!-\!\mathbf{p}')^2\;,
\label{eq:kinF}
\\
b^{(\pm)}_{00} &=& \frac{1}{16}\left[\pm\,(5+4x_2)\,t_2 + 3\,t_1\right]\;,
\label{defc00}\\
b^{(\pm)}_{10} &=& \frac{1}{16}\left[\pm\,(1+2x_2)\,t_2 - (1-2x_1)\,t_1\right]\;,
\label{defc10}\\
b^{(\pm)}_{01} &=& \frac{1}{16}\left[\pm\,(1+2x_2)\,t_2 - (1+2x_1)\,t_1\right]\;,
\label{defc01}\\
b^{(\pm)}_{11} &=& \frac{1}{16}\left[\pm\;t_2 - t_1\right]\;.
\label{defc11}
\end{eqnarray}
\end{subequations}
This is the full residual interaction as it must be taken into account
in a consistent Skyrme-RPA calculation.                The first line in Eq. (\ref{eq:kinF}) corresponds to the direct ($D$) term of
the interaction depicted in Fig. \ref{fig:basicV}, while the second line corresponds to
the exchange ($X$) term. 

Let us check what happens if one applies the Landau approximation
to this $F^{ph}_\mathrm{Sk,grad}$. 
It reads in the limit of nuclear matter
\be
\bfq = 0\quad,\quad \bfp^2 = \bfp'^2 = k_F^2\,.
\label{qpppnm}
\ee
In this case, we have
\be
  \bfq'^2 
  = 
  2k_F^2\bigl[\,1 - P_1(\cos\theta)\,\bigr]
  \quad,\quad
  \cos\theta 
  =
  \frac{\bfp\cdot\bfp'}{k_F^2}
  \quad,
\label{qp2nm}
\ee
where $k_F^2 = (3\pi^2\rho/2)^{2/3}$.
This wipes out totally the direct term $\propto\mathbf{q}^2$. It
remains (suppressing spin terms)
\begin{eqnarray}
 F^{ph}_\mathrm{Lan,grad}
 &=&
   2k_F^2\bigl[b^{(+)}_{00}1^{\sigma}1^{\tau} 
       +b^{(+)}_{01}1^{\sigma}(\bftau\bftau')\bigr]
   -2k_F^2\bigl[b^{(+)}_{00}1^{\sigma}1^{\tau} 
       +b^{(+)}_{01}1^{\sigma}(\bftau\bftau')\bigr]
    P_1(\cos\theta)\,\kappa_p
   \quad.
\label{eq:kinFLan}
\end{eqnarray}
The velocity dependent exchange terms contribute to the leading order
of F${^{ph}}$ {(f$_0$, f$'_0$) as well as to the next to leading
order (f$_1$, f$'_1$). The correct result is recovered for
$\kappa_p=1$. This factor was introduced for the purpose of analysis.
By varying $\kappa_p$ one can study the impact of the $p$-wave terms
($\propto P_1$) on the RPA results in finite nuclei.

To test the effect of the Landau approximation, we have computed the
the strength functions of the isoscalar giant monopole resonance
and of the electromagnetic $E1$ and $E2$ excitations in
$^{208}$Pb using the SHF parametrization SLy4 \cite{Cha98a} and SkT6
\cite{Ton84}. The two parametrizations have very different
momentum dependence, their effective masses are $m^*/m$ = 0.68
and $m^*/m$ = 1.0, respectively. The calculations were performed within
RPA in which the single-particle continuum was discretized using a
computational box of 18 fm radius.  The space of the single-particle
states was restricted to levels below 100~MeV. The full SHF residual
interaction as deduced from the Skyrme energy functional was used in
RPA.  Although it is known that spin-orbit and Coulomb terms are
crucial for a fully consistent Skyrme RPA calculation \cite{Sil06a},
we omit these terms here to allow a more direct comparison with the
residual interaction in the Landau approximation. The remaining part
of the residual interaction, including velocity-dependent terms, was
treated exactly in the calculations labeled by the symbol $D+X$ in the
figures.  The calculations indicated as $D=0$ employed the Landau
approximation (\ref{eq:kinFLan}) for the gradient term with different
values of $\kappa_p$ (in this case the direct term in Eq. (\ref{eq:kinF})
is completely omitted).
The effect of
$\mathbf{q}$-dependence in the full residual interaction
(\ref{eq:kinF}) is seen by comparison with the Landau approximation
(\ref{eq:kinFLan}) at $\kappa_p=1$. The impact of the $f_1$ and $f'_1$ terms is seen
by comparing different values of $\kappa_p$.
In the following, we discuss
three cases in detail.

\begin{figure}[!ht]
\begin{picture}(100,80)
\put(-10,-95){\includegraphics[scale=0.9,angle=0]{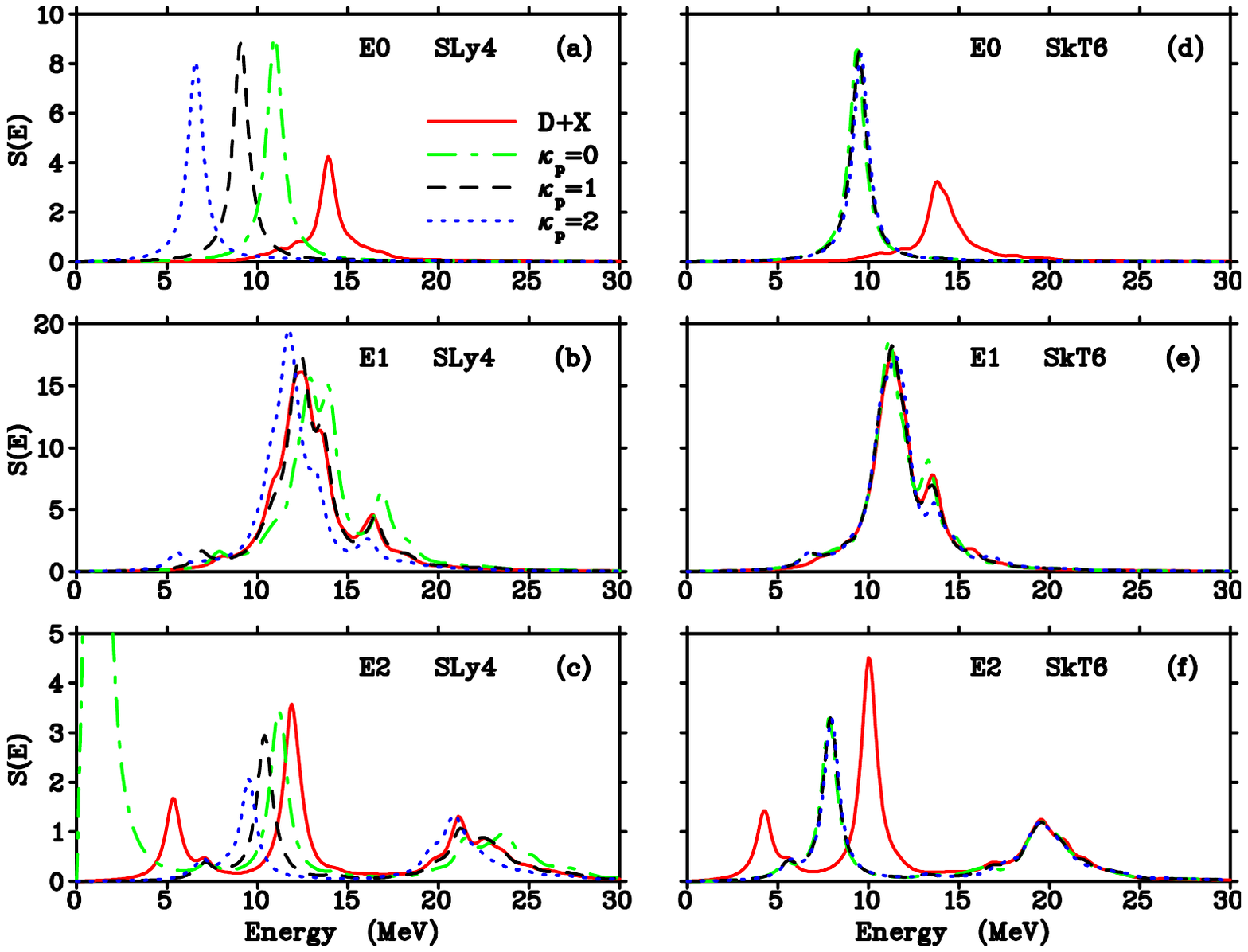}}
\end{picture}
\caption{\label{fig:1v}
Strength functions of the isoscalar $E0$ and electromagnetic
$E1$ and $E2$ excitations in $^{208}$Pb, calculated within the
self-consistent DFT+DRPA approach based on the Skyrme forces SLy4 \cite{Cha98a}
(panels (a), (b) and (c)) and SkT6 \cite{Ton84} (panels (d), (e) and (f)).
The discrete RPA spectra are folded with a Gaussian of width $\Delta=500$ keV.
The $E0$ and $E2$ strength functions are given in units
$10^3$fm$^4/$MeV, the $E1$ strength is given in units fm$^2/$MeV.
  The solid (red) lines represent the
  Skyrme RPA results using the full $ph$ interaction (\ref{eq:kinF}).
  The dashed (black), dotted (blue), and dashed-dotted (green) lines
  represent the results in Landau approximation ($D=0$, Eq. (\ref{eq:kinFLan}))
  for various values of the parameter $\kappa_p$ as indicated on the panel (a).
}
\end{figure}

Fig. \ref{fig:1v} shows that the $\mathbf{q}$-dependence of the residual
interaction is very important for the isoscalar resonances, see $E0$ and $E2$ distributions.
The energy of the first $2^+$ state in $^{208}$Pb becomes
even imaginary in the $D=0$ calculations with SLy4 force at $\kappa_p
= 1$ and $\kappa_p = 2$ and with SkT6 force at all $\kappa_p$.
The isoscalar $\mathbf{q}$-dependent part (parameter $b^{(-)}_{00}$ in Eq. (\ref{eq:kinF}))
is well defined by the ground state fits and varies little between
different parametrizations.  The effect of the isoscalar $p$-wave
from the kinetic terms differs very much between SLy4 and SkT6. SLy4
with the low $m^*/m$ shows a large dependence. SkT6, on the other
hand, reacts inert because $m^*/m=1$ means that there is no
isoscalar kinetic contribution ($b^{(+)}_{00}=0$).

The isovector giant dipole resonance (GDR) shown in the panels (b) and (e) of
Fig. \ref{fig:1v} is generally more robust. The
$\mathbf{q}$-dependence and the impact of the $p$-wave terms are much
smaller than for the isoscalar resonances. This relates to the
  fact that the isovector gradient terms are rather small for these
  two SHF parametrizations.  The effect of varying the $p$-wave
  contribution on the GDR is smaller than for the isoscalar modes
  which is due to a moderate TRK sum rule enhancement factor
  $\kappa_\mathrm{TRK}=0.25$ for the SLy4 force.
  The effect is even much smaller for the
  GDR with SkT6 force because this parametrization has
  vanishing $\kappa_\mathrm{TRK}$.

Altogether, we see that the Landau approximation can be
  disastrous in connection with Skyrme forces. It is not applicable in
  case of isoscalar modes for all relevant parametrizations. The case
  of isovector GDR is more forgiving. Parametrizations with high
  effective mass $m^*/m\approx 1$ and low TRK sum rule enhancement
  $\kappa_\mathrm{TRK}$ still allow to obtain acceptable results
  within the Landau approximation.

	\subsection{Landau-Migdal Interaction}
\label{secMigdal}

What do we learn from the present investigation on the Landau-Migdal interaction.
First of all one obtains in the isoscalar channel in next to leading order an attractive contribution depending on the magnitude of the effective mass. In heavy nuclei, where $m^*/m\approx 1$ this contribution is negligible. In medium mass and light nuclei where $m^*/m<1$ the effect could be of the order of one to two MeV. The consequence would be that f$_1$ should be A-dependent. From Eq. (\ref{eq:kinFLan}), however, one  notices that one obtains also a repulsive contribution to f$_0$  which makes the total effect of the exchange term slightly repulsive. This follows from the fact that for most of the known Skyrme parametrizations $b^{(\pm)}_{00} \geqslant 0$. In this respect the leading order is a good approximation. The ${\bf{q}}^2$-term which is neglected in the Landau approximation gives rise to a strong repulsion in the isoscalar channel, the magnitude of which is determined by the parameter $b^{(-)}_{00}$ which is of the same order for most of the Skyrme parametrizations. In particular, $b^{(-)}_{00}=125$ and 110 $\,\mbox{MeV}\cdot\mbox{fm}^5$ for SLy4 and SkT6 forces, respectively. This contribution is crucial in the self-consistent approach. In the phenomenological Landau-Migdal theory, however, this repulsion is included in the phenomenological parameters f$_0$. 

\section{Conclusion}

The Landau-Migdal theory is a theory for the excitation modes of
  a many-Fermion system. It can be derived microscopically from first
principles which leads eventually to an effective short-range
  interaction. For practical application, however, Landau's
quasi-particle concept is introduced which transforms the original
microscopic theory into a phenomenological approach. The central
quantity is the Landau-Migdal $ph$-interaction, which is
parametrized in terms of the famous Landau-Migdal
parameters. The Skyrme-Hartree-Fock (SHF) approach is a
self-consistent theory for ground-state and dynamics of
  nuclei. It also based on zero-range effective interactions whose
  parameters are determined by a fit to empirical data, mostly from
  the nuclear ground state. In this paper, we have used the well
  calibrated SHF approach to ``derive'' the Landau-Migdal parameters
  up to first order. Within this "microscopic" model we thus
  can test the Landau approximation and compare it with the
phenomenological results.

First, we have investigated the magnitude and density dependence
of the leading order (non-spin dependent) parameters deduced from
  SHF. The density dependence differs from the one of the
  Landau-Migdal parameters. However, the crucial values at densities
  corresponding to the nuclear surface region are in good agreement
  with the phenomenological Landau-Migdal parameters. The SHF results
  do also supply uncertainties on the predicted parameters. The
  smallest uncertainties are found again in the crucial surface
  region. SHF models which included information on excitation
  properties in their fit deliver very small uncertainties.

Second, we were able to consider for the first time the next to
leading order in the Landau-Migdal interaction. The attractive
contribution in the isoscalar channel, generated by the first
  order parameter f$_1$ parameter, is compensated by a repulsive
contribution to the leading order parameter f$_0$. Therefore the
leading order is sufficient for a phenomenological
  adjustment. The gradient terms in the SHF functional produce
  in the $ph$-interaction a strongly repulsive term proportional to
${\mathbf{q}}^2$ (where $\mathbf{q}$ is the transferred
  momentum) which is neglected in the Landau approximation. This
contribution is crucial in the Skyrme-Hartree-Fock approach. It
is implicitly considered in the phenomenological f$_0$ parameters.
The corresponding effects in the isovector channel are small and can
be neglected. Therefore we conclude: all derivative terms need to
  be carefully included in RPA based on SHF, but there is no need to
go beyond the leading order in the phenomenological
Landau-Migdal interaction.

\bigskip

\noindent
Acknowledgment: This work was supported by the Deutsche
Forschungsgemeinschaft (grant RE322-13/1). N. L. and V. T.
acknowledge financial support from the St. Petersburg State
University under Grant No. 11.38.648.2013. J.S. acknowledges 
travel support by the Alexander von Humboldt foundation.

\bibliography{Migdal}
\bibliographystyle{model1a-num-names}

\end{document}